\documentclass[aps,prl,preprint,tightenlines,superscriptaddress,showpacs,nofootinbib]{revtex4-1}
\usepackage[dvips]{graphicx}
\usepackage{epsfig}
\usepackage{dcolumn}
\usepackage{bm}
\usepackage{color}
\usepackage{ulem}

\newcommand{\gev}{\rm GeV}
\newcommand{\gevc}{{\rm GeV}/c}
\newcommand{\gevcs}{{\rm GeV}/c^2}
\newcommand{\mev}{\rm MeV}
\newcommand{\mevcs}{{\rm MeV}/c^2}

\newcommand{\ev}{\rm eV}

\newcommand{\MMS}{M_{\rm rec}^2}

\newcommand{\BR}{{\cal B}}

\newcommand{\chic}{\chi_{cJ}}

\newcommand{\psp}{\psi(2S)}

\newcommand{\jpsi}{J/\psi}

\newcommand{\EE}{e^+e^-}
\newcommand{\MM}{\mu^+\mu^-}

\newcommand{\ppjpsi}{\pi^+\pi^- J/\psi}

\newcommand{\pppsp}{\pi^+\pi^- \psp}
\newcommand{\pppsip}{\pi^+\pi^- \psp}

\newcommand{\reduline}{\bgroup\markoverwith
{\textcolor{red}{\rule[0.5ex]{2pt}{0.4pt}}}\ULon}

\newcommand{\beq}{\begin{equation}}
\newcommand{\eeq}{\end{equation}}
\newcommand{\bitm}{\begin{itemize}}
\newcommand{\eitm}{\end{itemize}}

\def\Journal#1#2#3#4{{#1} {\bf #2}, #3 (#4)}

\def\NIMA{Nucl. Instrum. Methods Phys. Res., Sect. A}

\def\PLB{Phys. Lett. B}
\def\PRL{Phys. Rev. Lett.}
\def\PRD{Phys. Rev. D}

\def\EPJC{Eur. Phys. J. C}

\begin{document}

%************************************************************
\preprint{
\vbox{ \hbox{   }
                        \hbox{Belle Preprint 2015-10}
                        \hbox{KEK   Preprint 2015-15}}
}
%\preprint{} \preprint{ \vbox{ \hbox{   }
%       \hbox{Intended for {\it Phys. Rev. D (R)}}
%        \hbox{Authors: Y. L. Han, X. L. Wang, C. Z. Yuan, C. P. Shen, P. Wang}
%        \hbox{Committee: Simon Eidelman (chair), Malachi Schram, Anatoly Sokolov}
%        }}
        
\title{
\quad\\[1.0cm]
Measurement of $\EE \to \gamma\chic$ via initial state radiation
at Belle}

%%% Paper:    e+ e- -> gamma chi_cJ
%%% Journal:  Physical Review D (Rapid Communication)
%%% Contacts: Y. Han (hanyl@ihep.ac.cn)
%%%           C. Z. Yuan (yuancz@ihep.ac.cn)
%%% Non-responding authors or those who said NO are commented out.
%%% ====================================================================
%%% Click the RELOAD button on your web browser to see the updated file.
%%% ====================================================================
%%% Use \input{author} to insert this material into your latex file.
%%%%% Force institutions to appear in alphabetical order when typeset.
\noaffiliation
\affiliation{University of the Basque Country UPV/EHU, 48080 Bilbao}
\affiliation{Beihang University, Beijing 100191}
%%%\affiliation{University of Bonn, 53115 Bonn}
\affiliation{Budker Institute of Nuclear Physics SB RAS, Novosibirsk 630090}
\affiliation{Faculty of Mathematics and Physics, Charles University, 121 16 Prague}
%%%\affiliation{Chiba University, Chiba 263-8522}
\affiliation{Chonnam National University, Kwangju 660-701}
\affiliation{University of Cincinnati, Cincinnati, Ohio 45221}
\affiliation{Deutsches Elektronen--Synchrotron, 22607 Hamburg}
%%%\affiliation{University of Florida, Gainesville, Florida 32611}
%%%\affiliation{Department of Physics, Fu Jen Catholic University, Taipei 24205}
\affiliation{Justus-Liebig-Universit\"at Gie\ss{}en, 35392 Gie\ss{}en}
\affiliation{Gifu University, Gifu 501-1193}
%%%\affiliation{II. Physikalisches Institut, Georg-August-Universit\"at G\"ottingen, 37073 G\"ottingen}
\affiliation{SOKENDAI (The Graduate University for Advanced Studies), Hayama 240-0193}
\affiliation{Gyeongsang National University, Chinju 660-701}
\affiliation{Hanyang University, Seoul 133-791}
\affiliation{University of Hawaii, Honolulu, Hawaii 96822}
\affiliation{High Energy Accelerator Research Organization (KEK), Tsukuba 305-0801}
%%%\affiliation{Hiroshima Institute of Technology, Hiroshima 731-5193}
\affiliation{IKERBASQUE, Basque Foundation for Science, 48013 Bilbao}
%%%\affiliation{University of Illinois at Urbana-Champaign, Urbana, Illinois 61801}
%%%\affiliation{Indian Institute of Technology Bhubaneswar, Satya Nagar 751007}
%%%\affiliation{Indian Institute of Technology Guwahati, Assam 781039}
\affiliation{Indian Institute of Technology Madras, Chennai 600036}
\affiliation{Indiana University, Bloomington, Indiana 47408}
\affiliation{Institute of High Energy Physics, Chinese Academy of Sciences, Beijing 100049}
\affiliation{Institute of High Energy Physics, Vienna 1050}
\affiliation{Institute for High Energy Physics, Protvino 142281}
%%%\affiliation{Institute of Mathematical Sciences, Chennai 600113}
\affiliation{INFN - Sezione di Torino, 10125 Torino}
\affiliation{Institute for Theoretical and Experimental Physics, Moscow 117218}
\affiliation{J. Stefan Institute, 1000 Ljubljana}
\affiliation{Kanagawa University, Yokohama 221-8686}
\affiliation{Institut f\"ur Experimentelle Kernphysik, Karlsruher Institut f\"ur Technologie, 76131 Karlsruhe}
%%%\affiliation{Kavli Institute for the Physics and Mathematics of the Universe (WPI), University of Tokyo, Kashiwa 277-8583}
\affiliation{Kennesaw State University, Kennesaw GA 30144}
\affiliation{King Abdulaziz City for Science and Technology, Riyadh 11442}
\affiliation{Department of Physics, Faculty of Science, King Abdulaziz University, Jeddah 21589}
\affiliation{Korea Institute of Science and Technology Information, Daejeon 305-806}
\affiliation{Korea University, Seoul 136-713}
%%%\affiliation{Kyoto University, Kyoto 606-8502}
\affiliation{Kyungpook National University, Daegu 702-701}
\affiliation{\'Ecole Polytechnique F\'ed\'erale de Lausanne (EPFL), Lausanne 1015}
\affiliation{Faculty of Mathematics and Physics, University of Ljubljana, 1000 Ljubljana}
%%%\affiliation{Ludwig Maximilians University, 80539 Munich}
\affiliation{Luther College, Decorah, Iowa 52101}
\affiliation{University of Maribor, 2000 Maribor}
\affiliation{Max-Planck-Institut f\"ur Physik, 80805 M\"unchen}
\affiliation{School of Physics, University of Melbourne, Victoria 3010}
\affiliation{Moscow Physical Engineering Institute, Moscow 115409}
\affiliation{Moscow Institute of Physics and Technology, Moscow Region 141700}
\affiliation{Graduate School of Science, Nagoya University, Nagoya 464-8602}
\affiliation{Kobayashi-Maskawa Institute, Nagoya University, Nagoya 464-8602}
%%%\affiliation{Nara University of Education, Nara 630-8528}
\affiliation{Nara Women's University, Nara 630-8506}
\affiliation{National Central University, Chung-li 32054}
%%%\affiliation{National United University, Miao Li 36003}
\affiliation{Department of Physics, National Taiwan University, Taipei 10617}
\affiliation{H. Niewodniczanski Institute of Nuclear Physics, Krakow 31-342}
%%%\affiliation{Nippon Dental University, Niigata 951-8580}
\affiliation{Niigata University, Niigata 950-2181}
\affiliation{University of Nova Gorica, 5000 Nova Gorica}
\affiliation{Novosibirsk State University, Novosibirsk 630090}
\affiliation{Osaka City University, Osaka 558-8585}
%%%\affiliation{Osaka University, Osaka 565-0871}
\affiliation{Pacific Northwest National Laboratory, Richland, Washington 99352}
%%%\affiliation{Panjab University, Chandigarh 160014}
\affiliation{Peking University, Beijing 100871}
%%%\affiliation{University of Pittsburgh, Pittsburgh, Pennsylvania 15260}
%%%\affiliation{Punjab Agricultural University, Ludhiana 141004}
%%%\affiliation{Research Center for Electron Photon Science, Tohoku University, Sendai 980-8578}
%%%\affiliation{Research Center for Nuclear Physics, Osaka University, Osaka 567-0047}
%%%\affiliation{RIKEN BNL Research Center, Upton, New York 11973}
%%%\affiliation{Saga University, Saga 840-8502}
\affiliation{University of Science and Technology of China, Hefei 230026}
%%%\affiliation{Seoul National University, Seoul 151-742}
%%%\affiliation{Shinshu University, Nagano 390-8621}
\affiliation{Soongsil University, Seoul 156-743}
\affiliation{University of South Carolina, Columbia, South Carolina 29208}
%%%\affiliation{Sungkyunkwan University, Suwon 440-746}
%%%\affiliation{School of Physics, University of Sydney, NSW 2006}
\affiliation{Department of Physics, Faculty of Science, University of Tabuk, Tabuk 71451}
\affiliation{Tata Institute of Fundamental Research, Mumbai 400005}
\affiliation{Excellence Cluster Universe, Technische Universit\"at M\"unchen, 85748 Garching}
\affiliation{Toho University, Funabashi 274-8510}
%%%\affiliation{Tohoku Gakuin University, Tagajo 985-8537}
\affiliation{Tohoku University, Sendai 980-8578}
\affiliation{Earthquake Research Institute, University of Tokyo, Tokyo 113-0032}
\affiliation{Department of Physics, University of Tokyo, Tokyo 113-0033}
\affiliation{Tokyo Institute of Technology, Tokyo 152-8550}
%%%\affiliation{Tokyo Metropolitan University, Tokyo 192-0397}
%%%\affiliation{Tokyo University of Agriculture and Technology, Tokyo 184-8588}
\affiliation{University of Torino, 10124 Torino}
%%%\affiliation{Toyama National College of Maritime Technology, Toyama 933-0293}
%%%\affiliation{Utkal University, Bhubaneswar 751004}
\affiliation{CNP, Virginia Polytechnic Institute and State University, Blacksburg, Virginia 24061}
\affiliation{Wayne State University, Detroit, Michigan 48202}
\affiliation{Yamagata University, Yamagata 990-8560}
\affiliation{Yonsei University, Seoul 120-749}
  \author{Y.~L.~Han}\affiliation{Institute of High Energy Physics, Chinese Academy of Sciences, Beijing 100049} % IHEP
  \author{X.~L.~Wang}\affiliation{CNP, Virginia Polytechnic Institute and State University, Blacksburg, Virginia 24061} % VPI
  \author{C.~Z.~Yuan}\affiliation{Institute of High Energy Physics, Chinese Academy of Sciences, Beijing 100049} % IHEP
  \author{C.~P.~Shen}\affiliation{Beihang University, Beijing 100191} % Beihang
  \author{P.~Wang}\affiliation{Institute of High Energy Physics, Chinese Academy of Sciences, Beijing 100049} % IHEP
  \author{A.~Abdesselam}\affiliation{Department of Physics, Faculty of Science, University of Tabuk, Tabuk 71451} % Tabuk
  \author{I.~Adachi}\affiliation{High Energy Accelerator Research Organization (KEK), Tsukuba 305-0801}\affiliation{SOKENDAI (The Graduate University for Advanced Studies), Hayama 240-0193} % KEK
% \author{K.~Adamczyk}\affiliation{H. Niewodniczanski Institute of Nuclear Physics, Krakow 31-342} % Krakow
  \author{H.~Aihara}\affiliation{Department of Physics, University of Tokyo, Tokyo 113-0033} % Tokyo
  \author{S.~Al~Said}\affiliation{Department of Physics, Faculty of Science, University of Tabuk, Tabuk 71451}\affiliation{Department of Physics, Faculty of Science, King Abdulaziz University, Jeddah 21589} % Tabuk
% \author{K.~Arinstein}\affiliation{Budker Institute of Nuclear Physics SB RAS, Novosibirsk 630090}\affiliation{Novosibirsk State University, Novosibirsk 630090} % BINP
% \author{Y.~Arita}\affiliation{Graduate School of Science, Nagoya University, Nagoya 464-8602} % Nagoya
  \author{D.~M.~Asner}\affiliation{Pacific Northwest National Laboratory, Richland, Washington 99352} % PNNL
% \author{T.~Aso}\affiliation{Toyama National College of Maritime Technology, Toyama 933-0293} % Toyama
% \author{V.~Aulchenko}\affiliation{Budker Institute of Nuclear Physics SB RAS, Novosibirsk 630090}\affiliation{Novosibirsk State University, Novosibirsk 630090} % BINP
% ...
% Ljubljana
% \author{N.~Zwahlen}\affiliation{\'Ecole Polytechnique F\'ed\'erale de Lausanne (EPFL), Lausanne 1015} % Lausanne
% \author{O.~Zyukova}\affiliation{Budker Institute of Nuclear Physics SB RAS, Novosibirsk 630090}\affiliation{Novosibirsk State University, Novosibirsk 630090} % BINP
\collaboration{The Belle Collaboration}

%\author{Belle Collaboration}

\date{\today}

\begin{abstract}
The process $\EE\to \gamma\chi_{cJ}$ ($J$=1, 2) is studied via
initial state radiation using 980~fb$^{-1}$ of data at and around
the $\Upsilon(nS)$ ($n$=1, 2, 3, 4, 5) resonances collected with
the Belle detector at the KEKB asymmetric-energy $\EE$ collider.
No significant signal is observed
except from $\psp$ decays. Upper limits on the cross
sections between $\sqrt{s}=3.80$ and $5.56~\gev$ are determined
at the 90\% credibility level,
which range from few pb to a few tens of pb.
We also set upper limits on the decay rate of the
vector charmonium [$\psi(4040$), $\psi(4160)$, and $\psi(4415)$] and
charmoniumlike [$Y(4260)$, $Y(4360)$, and $Y(4660)$] states to $\gamma\chi_{cJ}$.
%This information may help in understanding the nature of
%these vector states.

\end{abstract}


\maketitle

In $\EE$ annihilation, the energy region above the $D\overline{D}$
threshold is rich with vector charmonium and charmoniumlike
states. Three charmoniumlike states with $J^{PC}=1^{--}$ were
discovered at $B$ factories via initial state radiation (ISR) in
the last decade: the $Y(4260)$ in $\EE\to
\ppjpsi$~\cite{belley,babay4260} and the $Y(4360)$ and $Y(4660)$
in $\EE\to \pppsp$~\cite{pppsp,babay4324}. Together with the
conventional charmonium states $\psi(4040)$, $\psi(4160)$, and
$\psi(4415)$, there are six vector states; the potential
models predict only five in this mass
region~\cite{barnes}. Some of these states show
unusual properties that are inconsistent with charmonium~\cite{review}.
It is unlikely that all of these states are
charmonia; some, perhaps,  have exotic nature: a multiquark
state, molecule, hybrid, or some other configuration. To improve our
understanding of these states and the underlying QCD, it is important to investigate them using
much larger data samples and new decay channels.

For example, one can study radiative
transitions between these states and lower charmonium states like
the $\chic$.
%The only existing investigation of such a type is from the
%CLEO Collaboration which reported $\sigma(\EE\to
%\gamma\chic)<126$~pb at center-of-mass (CM) energy $\sqrt{s} =
%3970-4060~\mev$, $\sigma(\EE\to \gamma\chic)<79$~pb at $\sqrt{s} =
%4120-4200~\mev$, and $\sigma(\EE\to \gamma\chic)<120$~pb at
%$\sqrt{s} =4260~\mev$~\cite{cleo_br}, at the 90\% confidence level
%(C.L.), with the scan data between 3.97 and 4.26~GeV.
%CLEO Collaborationhas done a such type investigation,
 %which used data taken during a scan of center-of-mass (CM)
%energies $\sqrt{s}=3.97-4.26~\gev$ to report upper limits on the cross
%sections of $\EE \to \gamma\chi_{c1}$ and  $\EE \to \gamma\chi_{c2}$
%in three energy regions: the $\psi(4040)$ ($\sqrt{s}$ = 3.97-4.06~GeV),
%the $\psi(4160)$ (4.12-4.20~GeV), and $\sqrt{s}$ =
%4.26~GeV.
%The limited statistics of the CLEO analysis prevented
%them from measuring the line shape of $\EE\to \gamma\chic$.
%BES\uppercase\expandafter{\romannumeral3} experiment has also done a scan
%at four energy points: $\sqrt{s}=$ 4.009, 4.230, 4.260, and 4.360~GeV,
%and reports the upper limits on the cross section of
%$\EE \to \gamma\chi_{c1}$ and  $\EE \to \gamma\chi_{c2}$ at these energy
%points~\cite{cepc}.
%With the full Belle data sample, we can study this process via
%ISR, and will have better sensitivity than in the CLEO measurement.
The CLEO Collaboration used data taken during a scan of center-of-mass (CM) 
energies $\sqrt{s} = 3.97-4.26~\gev$ to report upper limits on the
cross sections of $\EE \to \gamma\chi_{c1}$ and  $\EE \to \gamma\chi_{c2}$
in three energy regions: the $\psi(4040)$ ($\sqrt{s}$ = 3.97-4.06~GeV),
the $\psi(4160)$ (4.12-4.20~GeV), and $\sqrt{s}$ =
4.26~GeV~\cite{cleo_br} .
The limited statistics prevented them from measuring 
the line shape of  $\EE\to \gamma\chic$. The BESIII experiment reports the
 upper limits on the cross sections of
the reactions  $\EE \to \gamma\chi_{c1}$ and  $\EE \to \gamma\chi_{c2}$
 at four energy points: $\sqrt{s}$ = 4.009, 4.230, 4.260, and 4.360 GeV~\cite{bes3}.
With the full Belle data sample, we are able to study this process via ISR.

In this paper, we report a study of the $\EE \to \gamma\chic$ process
using ISR events detected with the Belle detector~\cite{Belle} at
the KEKB asymmetric-energy $\EE$ collider~\cite{KEKB}. Here,
$\chic$ is reconstructed in the $\gamma\jpsi$ final state and $\jpsi$ is
reconstructed in the $\MM$ final state alone (The background level is
very high in the $\EE$ final state due to Bhabha events). The same final state 
 $\gamma \gamma J/\psi$, has been previously 
analyzed at Belle and $\psi(4040)$ and $\psi(4160)$
were observed as $\eta J/\psi$ resonances~\cite{etaj}. 
%The integrated luminosity used in this analysis is 980~fb$^{-1}$. 
We study the full Belle dataset corresponding to an integrated luminosity of
980~fb$^{-1}$. 
About 70\% of the data were collected
at the $\Upsilon(4S)$ resonance, and the remainder were taken at the other
$\Upsilon(nS)$ ($n$=1, 2, 3, or 5) states or at CM energies a few
tens of $\mev$ lower than the $\Upsilon(4S)$ or the $\Upsilon(nS)$
peaks.

The event generator {\sc evtgen}~\cite{EvtGen} with the {\sc
vectorisr} model is used to simulate the signal process $\EE \to
\gamma_{\rm ISR} V\to \gamma_{\rm ISR}\gamma\chic \to \gamma_{\rm
ISR}\gamma\gamma\jpsi$. The mass and width of $V$ can be varied
so that we can obtain the signal efficiency as a function of the
vector meson mass. This model  considers the
leading-order (LO) quantum electrodynamics (QED) correction  only and 
%only one photon is allowed to be emitted, 
thus higher-order corrections should be
estimated and properly  taken into account. The dedicated ISR generator {\sc
phokhara}~\cite{phokhara} has the next-to-leading-order (NLO) QED
correction but does not contain the mode of interest.
However, the process $\EE \to \gamma_{\rm ISR} V \to \gamma_{\rm
ISR} \eta\jpsi$ can be generated with {\sc phokhara} and this allows
us to estimate the NLO correction effect in the mode under study by
comparing the results from the two generators in the analysis of the
$\eta\jpsi$ mode.
All generated events are passed through the GEANT3~\cite{geant4} 
based detector simulation and then the standard reconstruction.

For a candidate event, we require two good charged tracks with
zero net charge.
%A good charged track has impact parameters with
%respect to the interaction point of $|dr| < 0.5$~cm in the
%$r$-$\phi$ plane and $|dz|<5$~cm in the $r$-$z$ plane.
The impact parameters of these tracks
perpendicular to and along the beam direction with respect to the
interaction point are required to
be less than 0.5 cm and 5.0 cm, respectively.
The transverse momentum of the leptons is required to be greater than
$0.1~\gevc$.
For each charged track, information from different detector subsystems is
combined to form a likelihood ${\cal L}_i$  for each particle species ($i$)~\cite{Rlike}.
%$R_{\mu}$ is defined as $R_{\mu} = \frac{\cal{L_{\mu}}}{\cal{L_{\mu}} + \cal{L_{\pi}}}$.
For muons from $\jpsi \to \MM$,
one of the tracks is required to have the
muon identification likelihood ratio
$\mathcal{R}_{\mu} = \frac{\cal{L_{\mu}}}{\cal{L_{\mu}} + \cal{L_{\pi}}} > 0.95$;
in addition, if one of the muon
candidates has no muon identification (ID)
information~\cite{MUID}, the polar angle of each muon candidate in the
$\gamma\chic$ CM system is required to satisfy
$|\cos\theta_{\mu}|< 0.75$. The lepton ID efficiency is about 87\%
for $\jpsi\to \MM$.

\begin{figure}[htb]
 \psfig{file=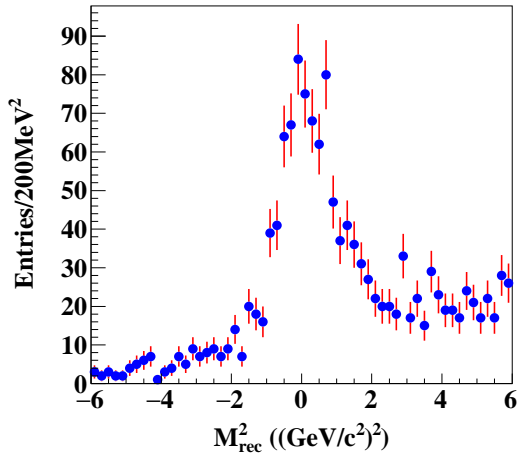, width=8cm}
\caption{Missing mass squared distribution with $M(\gamma_l\gamma_h\jpsi)<5.56$~GeV/$c^2$. } 
\label{mms}
\end{figure}

A photon candidate is an electromagnetic
calorimeter cluster with energy $E(\gamma) >
50~\mev$ that does not match any charged tracks. The photon
is labeled as the ISR photon when its energy in the $\EE$ CM
frame exceeds 3~$\gev$ (corresponding to $M[\gamma\chi_{cJ}]<7~\gevcs$, the maximum
non-ISR photon energy being about $3~\gev$) and this photon is excluded when reconstructing
$\gamma\chi_{cJ}$ candidates. 
%We require that there to be at
%least two additional photons and the two highest energy photons in
%the laboratory system (excluding the ISR photon) are selected,
%which are denoted as $\gamma_{h}$ and $\gamma_{l}$ ($E_{\gamma_{h}} > E_{\gamma_{l}}$).
%The energies of the photon candidates in the laboratory frame are
%required to satisfy $E(\gamma_{l}) > 0.25~\gev$. 
We also require at least two additional photons, each with energy in the laboratory frame 
greater than $0.25~\gev$. Among these, we select the two with the highest energy in the
laboratory system and denote these as
$\gamma_{h}$ and $\gamma_{l}$ (with $E_{\gamma_{h}} > E_{\gamma_{l}}$).
The detection of
the ISR photon is not required; instead, we require $-1~(\gevcs)^2
< \MMS < 2~(\gevcs)^2$, where $\MMS$ is the square of the mass
recoiling against the $\gamma\chic$ system.  The distribution of
$\MMS$  is shown in Fig.~\ref{mms}.

\begin{figure}[htb]
 \psfig{file=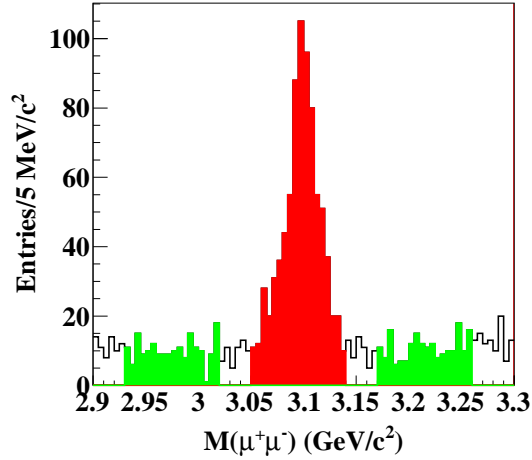, width=8cm}
\caption{Invariant mass distribution of $\MM$. The shaded area
in the middle is the $\jpsi$ signal region, and the shaded regions
on either side are the $\jpsi$ mass sidebands.} \label{mll}
\end{figure}

Fig.~\ref{mll} shows the $\MM$ invariant mass [$M({\MM})$] distribution for events that
survive the selection criteria and with the
$\gamma_l\gamma_h\jpsi$ invariant mass [$M(\gamma_l\gamma_h\jpsi)
= M(\gamma_l\gamma_h\MM) - M(\MM) + m_{J/\psi}$] less than
5.56~GeV/$c^2$, where $m_{J/\psi}$ is the nominal mass of the $\jpsi$~\cite{PDG}.
 A $\MM$ pair is
considered as a $\jpsi$ candidate if $M({\MM})$ is within $\pm
45~\mevcs$ (the mass resolution being $15~\mevcs$) of the $\jpsi$
nominal mass~\cite{PDG}.
The $\jpsi$ mass sidebands are defined as
$M({\MM})\in [3.172,~3.262]~\gevcs$ or $[2.932,~3.022]~\gevcs$,
which are twice as wide as the signal region.

\begin{figure}[htb]
 \psfig{file=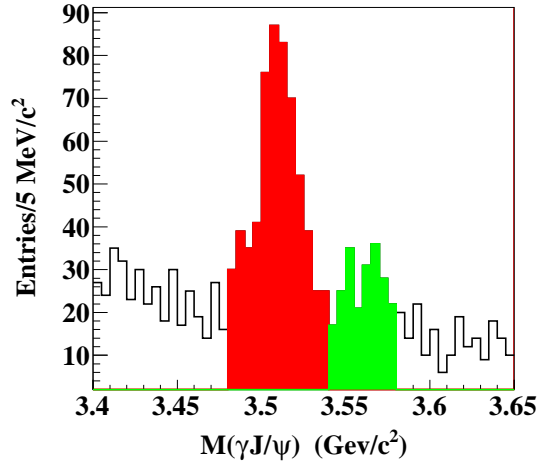,width=8cm}
\caption{Invariant mass distribution of $\gamma\jpsi$ for
candidate events with $M(\gamma_l\gamma_h\jpsi)<5.56$~GeV/$c^2$.
The shaded histograms show the $\chi_{c1}~([3.48,~3.535]~\gevcs)$
and $\chi_{c2}~([3.535,~3.58]~\gevcs)$ regions.}
\label{mchic_combine_total}
\end{figure}

To reject the background from $\EE \to \gamma_{\rm ISR} \eta(\pi^{0}) \jpsi$ events
with $\eta$ or $\pi^{0}$ decaying into two photons,
we require that the invariant mass of
the two photons, $M(\gamma\gamma)$, be outside the  $\eta$ mass region of
$[0.50,~0.58]~\gevcs$,  the $\pi^{0}$ mass region and the low-invariant-mass
region $M(\gamma\gamma)<0.20~\gevcs$.
Figure~\ref{mchic_combine_total} shows the invariant mass
distribution of $M(\gamma\jpsi)$ (with two entries per event for $M(\gamma_{h}\jpsi)$
and $M(\gamma_{l}\jpsi)$) for events with
$M(\gamma_l\gamma_h\jpsi)<5.56$~GeV/$c^2$. Here,
$M(\gamma_{l(h)}\jpsi) = M(\gamma_{l(h)}\MM)-M(\MM)+m_{J/\psi}$. We observe
 $\chi_{c1}$ and $\chi_{c2}$ signals but no evidence of $\chi_{c0}$. We divide the
$\chi_{cJ}$ mass region into $[3.48,~3.535]~\gevcs$ for $\chi_{c1}$
and $[3.535,~3.58]~\gevcs$ for $\chi_{c2}$.

\begin{figure}[htb]
 \psfig{file=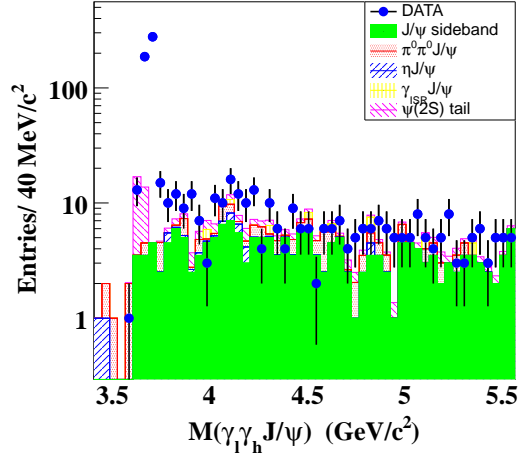,width=8cm}
\caption{Invariant mass distribution of
$\gamma_{l}\gamma_{h}\jpsi$.
The background from the tail of the
$\psp$ is plotted only for $M(\gamma_{l}\gamma_{h}\jpsi) >
3.75~\gevcs$ and $M(\gamma_{l}\gamma_{h}\jpsi) < 3.65~\gevcs$.
The dots with error bars are data while the shaded histograms
represent different sources of background modes.
}
\label{mgchc_full}
\end{figure}

Figure~\ref{mgchc_full} shows the $M(\gamma_{l}\gamma_{h}\jpsi)$
distribution after applying all the selection criteria above.
We  see a clear $\psp$ signal but no
significant signal in the higher mass region.
%although there are hints at some energies. 
The clear $\chi_{cJ}$ and $\psp$ signals allow
us to measure the product branching fractions $\BR[\psp\to
\gamma\chi_{cJ}] \times \BR[\chi_{cJ} \to \gamma\jpsi]$ $(J=1,~2)$.
By contrast, in the region $M(\gamma\gamma\jpsi) \in [3.80,~5.56]~\gevcs$,
we set an upper limit on the
production cross section of $\EE \to \gamma\chi_{cJ}$.

The potential backgrounds are also shown in Fig.~\ref{mgchc_full}.
Besides the non-$\jpsi$ background, which also appear in the $\jpsi$ mass sidebands, 
there are three additional
backgrounds: $\EE \to \gamma_{\rm ISR}\jpsi$,
$\gamma_{\rm ISR} \pi^{0}\pi^{0}\jpsi$, and $\gamma_{\rm ISR}
\eta\jpsi$. Of course, $\EE \to \gamma_{\rm ISR}\psp$ with
$\psp\to \gamma\chic$ will be a background in the analysis of the
$\gamma_l\gamma_h\jpsi$ high-mass region. The ISR $\jpsi$ and $\psp$
samples are generated according to the theoretical calculation of
the production cross sections~\cite{ISR} with the world-average resonant
parameters as input~\cite{PDG}. For the other modes, we use
the cross sections of $\EE \to \eta\jpsi$~\cite{etaj} and $\EE
\to \pi^{+}\pi^{-}\jpsi$~\cite{ppjpsi} and assume that
$\sigma(\EE \to \pi^{0}\pi^{0}\jpsi) =
\frac{1}{2}\sigma(\EE \to \pi^{+}\pi^{-}\jpsi)$. All these samples are generated
using the {\sc phokhara} generator~\cite{phokhara} and are normalized
to the integrated luminosity of the full data sample. The
background contribution practically saturates the mass spectrum above
the $\psp$ peak.

\begin{figure}[htb]
 \psfig{file=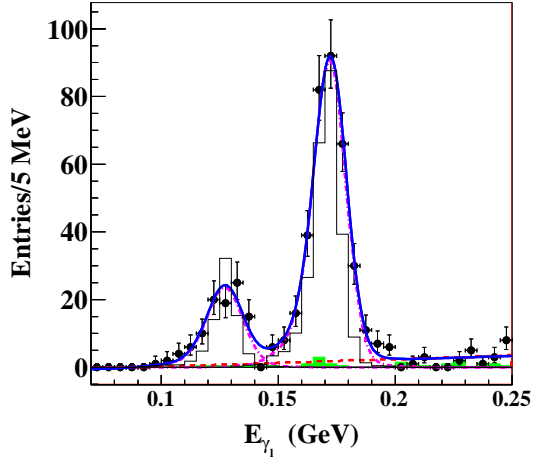,width=8cm}
\caption{
Energy distributions of the low energy photon in the
$\gamma_l\gamma_h\jpsi$ CM system for events in the $\psp$ mass
region. Dots with error bars are data and histograms are MC
samples. The blue solid line is the best fit,
the red dashed line is the shape of the total background determined from the fit, and
the purple dot-dashed line is the MC signal shape convolved with a Gaussian
function. The shaded histogram shows the total background as determined 
from $\jpsi$ sidebands and simulations.}

\label{fit_egl}
\end{figure}

To measure the $\psp\to \gamma\chi_{cJ}$ branching
fractions, we define the $\psp$ signal
region as $3.65~\gevcs<M(\gamma_l\gamma_h\jpsi)<3.72~\gevcs$.
The distribution of the energy of the less energetic photon in
the $\gamma_l\gamma_h\jpsi$ CM system is shown in
Fig.~\ref{fit_egl}. Clear signals due to $\chi_{c1}$ and $\chi_{c2}$ are
observed with very low background and 
%we fit Fig.~\ref{fit_egl}
we fit this photon energy distribution
to extract the corresponding yields. The $\chi_{cJ}$ signal shapes are
obtained from Monte Carlo simulated signal samples convolved with a
corresponding smearing Gaussian function to compensate for the
resolution difference between data and Monte Carlo simulation; the
background is parameterized as a first-order Chebyshev polynomial.
%The fitting results are shown in Fig.~\ref{fit_egl}, and the fit
%yields $1261 \pm 38$ and $619 \pm 29$ $\chi_{c1}$ and $\chi_{c2}$ signal events,
%respectively.
The resulting fit function is shown in Fig.~\ref{fit_egl} and the fit yields $340 \pm 20$ $\chi_{c1}$
and $97\pm 12$  $\chi_{c2}$ signal events.

From the world-average $\psp$ resonant parameters~\cite{PDG},
we calculate $\sigma[\EE \to \gamma_{\rm ISR} \psp]$ = $(14.25\pm
0.26)$~pb~\cite{ISR}
and thus expect $13.9 \times 10^6$
ISR produced $\psp$ events in the full Belle data sample  of 980 fb$^{-1}$.
With the efficiencies of 1.4\% and 0.7\% for the $\chi_{c1}$ and
$\chi_{c2}$ modes, respectively, from the MC simulation, we obtain
 $\BR[\psp\to \gamma\chi_{c1}] \times \BR(\chi_{c1} \to \gamma\jpsi)
 = (2.92 \pm 0.19)\%$ and
 $\BR[\psp\to \gamma\chi_{c2}] \times \BR(\chi_{c2} \to \gamma\jpsi)
 = (1.65 \pm 0.21)\%$.
Here, the errors are statistical only. These results are consistent
with the PDG values~\cite{PDG}.

\begin{figure}[htbp]
 \psfig{file=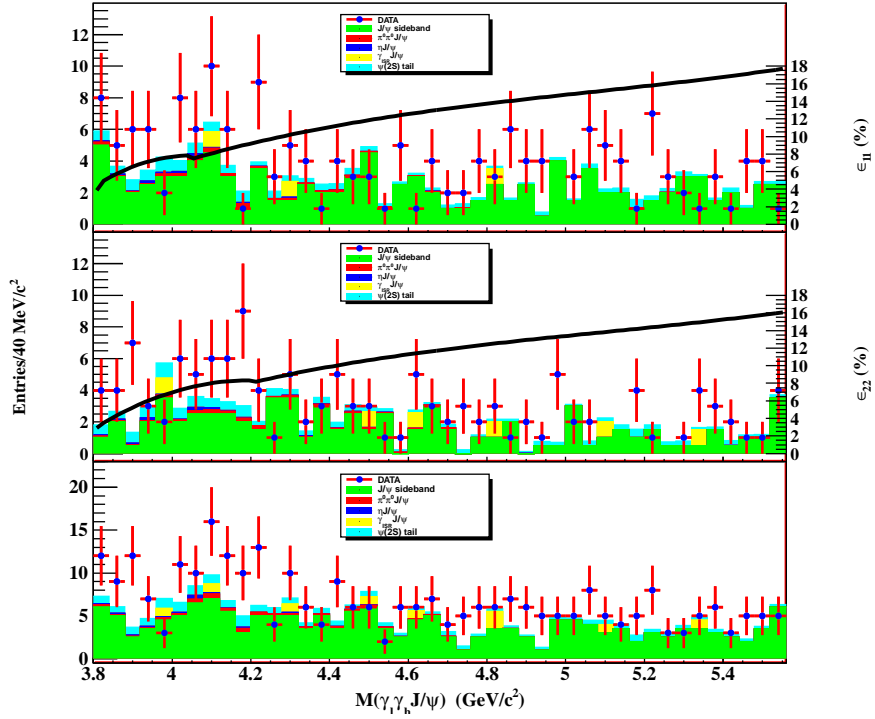,width=0.8\columnwidth}
\caption{Invariant mass distributions of $\gamma\chi_{cJ}$
candidates. Shown from top to bottom are $\gamma\chi_{c1}$,
$\gamma\chi_{c2}$, and their sum. Dots with error bars are data, the shaded histograms
are the simulated backgrounds and $\jpsi$ sidebands, and the solid lines are the
efficiency curves. } \label{mgchc_high_comp}
\end{figure}

The $M(\gamma_l\gamma_h\jpsi)$ distributions above the $\psi(2S)$ signal region for $\gamma\chi_{c1}$
and $\gamma\chi_{c2}$ candidate events as well as their sum
are shown in Fig.~\ref{mgchc_high_comp}, 
together with the background estimation from the $\jpsi$
mass sidebands and the MC simulated background modes with a
genuine $\jpsi$. No significant signal is observed in either
the $\gamma\chi_{c1}$ or $\gamma\chi_{c2}$ mode. As the background
estimation is limited to the known channels, it only serves as a
lower limit of the true background. In calculating the upper
limits of the $\gamma\chi_{cJ}$ production cross section, we
consider the estimated-background events from the observed signal
candidates. 
%This results in a conservative estimate of the upper
%limit of the signal events, so the upper limit of the cross section determined this way
%is conservative.
This results in a conservative estimate of the upper limit of the signal
and hence a conservative estimate for the cross section.

There is cross contamination between the $\chi_{c1}$ and $\chi_{c2}$
signals due to the mass resolution, as can be seen from
Fig.~\ref{mchic_combine_total}, and this is taken into account
as follows. The yields of observed $\chi_{c1}$ and $\chi_{c2}$
events (denoted as $n^{\chi_{c1}}_{\rm obs}$ and
$n^{\chi_{c2}}_{\rm obs}$, respectively) are expressed as

\begin{equation}
\left( \begin{array}{c}n^{\chi_{c1}}_{\rm obs} \\ n^{\chi_{c2}}_{\rm obs}\end{array}\right) =
\left(\begin{array}{cc}\epsilon_{11} &\epsilon_{21} \\ \epsilon_{12} &\epsilon_{22} \end{array}\right)
\left( \begin{array}{c}N^{\chi_{c1}} \times \BR(\chi_{c1} \to \gamma\jpsi) \times \BR(\jpsi \to \MM) \\
 N^{\chi_{c2}} \times \BR(\chi_{c2} \to \gamma\jpsi) \times \BR(\jpsi \to \MM)\end{array}\right)
 + \left( \begin{array}{c}n^{\chi_{c1}}_{\rm bkg} \\ n^{\chi_{c2}}_{\rm bkg}\end{array}\right).
\end{equation}

In these equations, $\epsilon_{ij}$ $(i,~j=1,~2)$ is the
efficiency of produced $\chi_{ci}$ to be reconstructed in the
$\chi_{cj}$ signal region; $N^{\chi_{c1}}$ and $N^{\chi_{c2}}$
represent the total numbers of $\chi_{c1}$ and $\chi_{c2}$ events produced in
data, respectively; $\BR$ is the world-average  branching
fraction for the given process~\cite{PDG}; and $n^{\chi_{c1}}_{\rm bkg}$
and $n^{\chi_{c2}}_{\rm bkg}$ represent the numbers of
non-$\chi_{cJ}$ background events for $\chi_{c1}$ and $\chi_{c2}$,
respectively, which are the sum of the normalized $\jpsi$ mass
sideband background and the MC simulated $\gamma_{\rm ISR}\jpsi$,
$\gamma_{\rm ISR}\eta\jpsi$, $\gamma_{\rm ISR}\pi^0\pi^{0}\jpsi$,
and $\gamma_{\rm ISR}\psp$ background, as shown in
Fig.~\ref{mgchc_high_comp}. The efficiency curves $\epsilon_{11}$
and $\epsilon_{22}$, also shown in Fig.~\ref{mgchc_high_comp}, are
not monotonic between
$3.9~\gevcs<m(\gamma\chi_{cJ})<4.2~\gevcs$. This is due to the fact that
the energies of the two photons are almost the same in this mass
region.

We use the maximum likelihood method to determine upper limits on
the numbers of produced $\gamma\chi_{cJ}$ events, $N^{\chi_{c1}}$
and $N^{\chi_{c2}}$ and thus on the upper limits of the production cross sections of
$\EE\to \gamma \chi_{cJ}$. The likelihood is constructed as
follows. For each possible pair  of the $N^{\chi_{c1}}$ and
$N^{\chi_{c2}}$ values, the numbers of the expected signal events, 
$\nu^{\chi_{c1}}$ and $\nu^{\chi_{c2}}$, are
\begin{equation}
 \label{nunu}
\left( \begin{array}{c}\nu^{\chi_{c1}} \\ \nu^{\chi_{c2}}
\end{array}\right) = \left(\begin{array}{cc}\epsilon_{11}
&\epsilon_{21} \\ \epsilon_{12} &\epsilon_{22} \end{array}\right)
\left( \begin{array}{c}N^{\chi_{c1}} \times \BR(\chi_{c1} \to \gamma\jpsi) \times \BR(\jpsi \to \MM) \\
 N^{\chi_{c2}} \times \BR(\chi_{c2} \to \gamma\jpsi) \times \BR(\jpsi \to
 \MM)\end{array}\right).
\end{equation}
Taking into account the background contribution, the numbers of
expected events in the signal regions, denoted as
$\mu^{\chi_{c1}}$ and $\mu^{\chi_{c2}}$ for $\chi_{c1}$ and
$\chi_{c2}$, respectively, are
\begin{equation}
 \label{mumu}
 \left( \begin{array}{c}\mu^{\chi_{c1}} \\ \mu^{\chi_{c2}} \end{array}\right) =
 \left( \begin{array}{c}\nu^{\chi_{c1}} \\ \nu^{\chi_{c2}} \end{array}\right)
 + \left( \begin{array}{c}n^{\chi_{c1}}_{\rm bkg} \\ n^{\chi_{c2}}_{\rm
 bkg}\end{array}\right),
\end{equation}
and the probability of observing
 $\left(^{n^{\chi_{c1}}_{\rm obs}}_{n^{\chi_{c2}}_{\rm obs}}\right)$
events in data is
\begin{equation}
 \label{p}
p(N^{\chi_{c1}}, N^{\chi_{c2}}) =
\frac{(\mu^{\chi_{c1}})^{n^{\chi_{c1}}_{\rm
obs}}e^{-\mu^{\chi_{c1}}}} {n^{\chi_{c1}}_{\rm obs}!}
\frac{(\mu^{\chi_{c2}})^{n^{\chi_{c2}}_{\rm
obs}}e^{-\mu^{\chi_{c2}}}} {n^{\chi_{c2}}_{\rm obs}!}.
\end{equation}
The uncertainty in the background estimation is considered by
sampling $n^{\chi_{cJ}}_{\rm bkg}$ in Eq.~(\ref{mumu}).
By fitting the normalized background distribution,
the mean value and the uncertainty of the background level are obtained.
The background yield  $n_{\rm bkg}^{\chi_{cJ}}$ is varied assuming it follows a Gaussian distribution
with this mean value and the uncertainty as the standard deviation.
The systematic error of the measurement, which corresponds to an
uncertainty in the expected number of events, follows a Gaussian
distribution with a mean value $\nu^{\chi_{cJ}}$ and a standard
deviation $\nu^{\chi_{cJ}}\times\sigma_{\rm sys}$, where
$\sigma_{\rm sys}$ is the total relative systematic error (13.4\%),
described below. This is also considered by varying
$\mu^{\chi_{cJ}}$ in Eq.~(\ref{p}).

\begin{figure}[htbp]
 \psfig{file=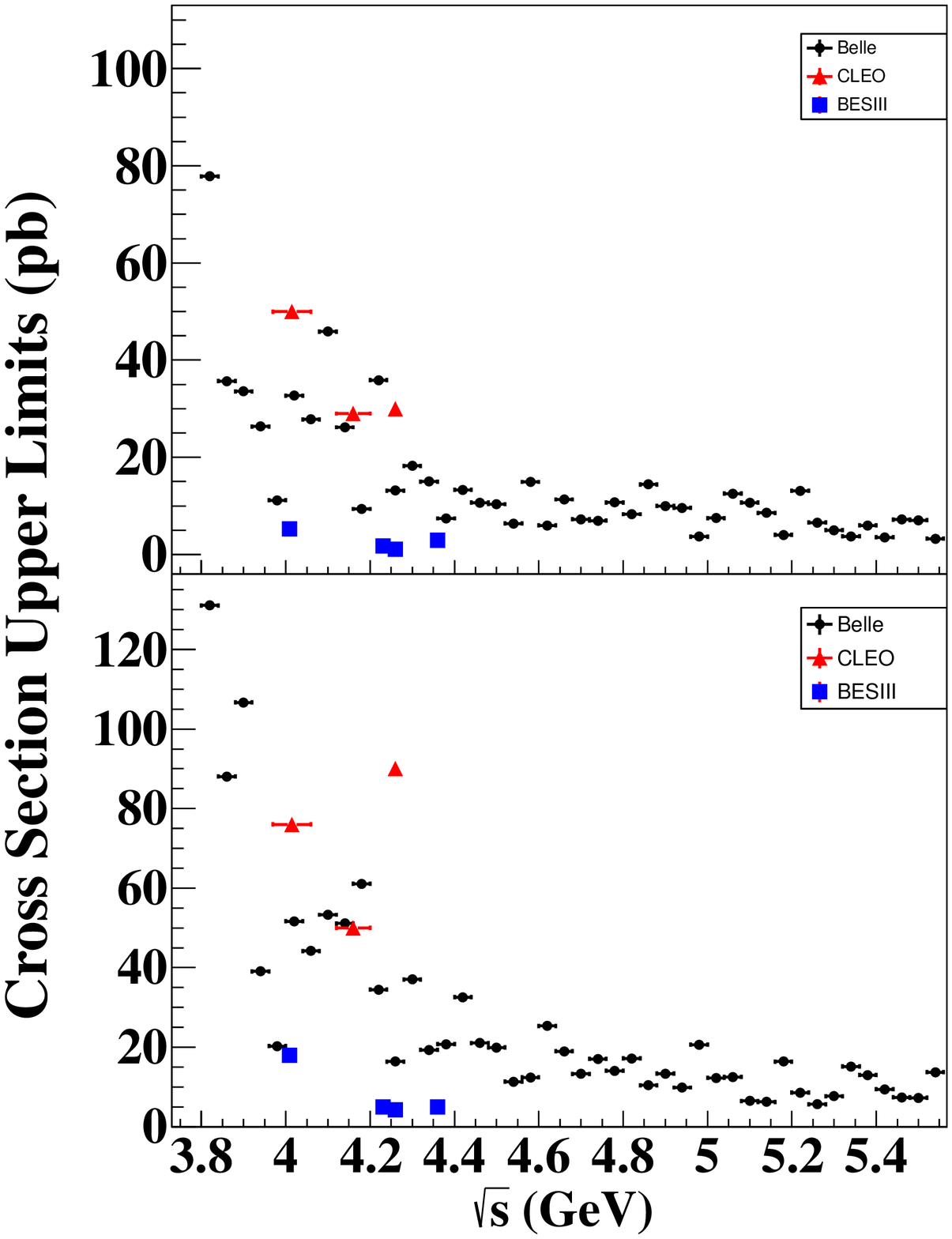, width=0.8\columnwidth}
\caption{Measured upper limits on the $\EE\to \gamma\chi_{cJ}$
cross sections at the 90\% C.L. for $\chi_{c1}$ (top) and $\chi_{c2}$
(bottom). The solid dots show the
Belle measurements, the solid triangles are the results from
CLEO and the blue squares are from BESIII.} \label{up_limit}
\end{figure}

The summation of random-sampled 
%The weighted 
$p(N^{\chi_{c1}}, N^{\chi_{c2}})$, considering
the uncertainty in background estimation and the systematic errors,
forms the final likelihood function 
%$L(N^{\chi_{c1}}, N^{\chi_{c2}})$ in Eq.~(\ref{likehood}).

\begin{equation}
 \label{likehood}
L(N^{\chi_{c1}}, N^{\chi_{c2}}) = \frac{1}{N}
\sum_{k,l,m,n}p(N^{\chi_{c1}}, N^{\chi_{c2}}) = \frac{1}{N}
\sum_{k,l,m,n}\frac{(\mu^{\chi_{c1}}_{k,l})^{n^{\chi_{c1}}_{\rm
obs}}e^{-\mu^{\chi_{c1}}_{k,l}}} {n^{\chi_{c1}}_{\rm obs}!}
\frac{(\mu^{\chi_{c2}}_{m,n})^{n^{\chi_{c2}}_{\rm
obs}}e^{-\mu^{\chi_{c2}}_{m,n}}} {n^{\chi_{c2}}_{\rm obs}!}.
\end{equation}
Here, $N$ is the number of samplings.
$\mu^{\chi_{c1}}_{k,l} = \nu^{\chi_{c1}}_{k} + n^{\chi_{c1}}_{{\rm bkg},l}$
and $\mu^{\chi_{c2}}_{m,n}= \nu^{\chi_{c2}}_{m} +n^{\chi_{c2}}_{{\rm bkg},n}$,
where $\nu^{\chi_{c1}}_{k}$,
$n^{\chi_{c1}}_{{\rm bkg},l}$, $\nu^{\chi_{c2}}_{m}$ and
$n^{\chi_{c2}}_{{\rm bkg},n}$ are the numbers of events obtained from
the corresponding Gaussian distributions.
The subscript $k$ represents the $k$-th sampling for the 
expected number of $\chi_{c1}$ signal events $\nu^{\chi_{c1}}$.
The other subscripts $l$, $m$ and $n$ have parallel meanings.
By letting $N^{\chi_{c1}}$ and $N^{\chi_{c2}}$ run over
all the possible values from 0 to infinity independently, we obtain
the likelihood in the $(N^{\chi_{c1}}, N^{\chi_{c2}})$ plane.
The likelihood $L(N^{\chi_{c1}})$ can be obtained from this two-dimensional
likelihood function by  integrating over the variable $N^{\chi_{c2}}$.
From this,  we obtain the upper limit on $N^{\chi_{c1}}$
at the 90\% credibility level (C.L.) 
\footnote{In common high energy physics usage, this
Bayesian interval has been reported as ``confidence interval'' which
is a frequentist-statistics term.}
and convert this into the upper limit on $\sigma(\EE \to \gamma\chi_{c1})$.
The upper limit on $\sigma(\EE \to \gamma\chi_{c2})$ is determined in a similar manner.
The final upper limits are shown in Fig.~\ref{up_limit} and are around 
a few pb to a few tens of pb. We also show the CLEO and BESIII
results in Fig.~\ref{up_limit} for comparison.  The measured upper
limits are more stringent than the CLEO results at
$\sqrt{s}=3.97-4.06~\gev$ and $\sqrt{s}=4.26~\gev$. 
The large data samples collected by BESIII at 
$\sqrt{s}=$ 4.009, 4.230, 4.260, and 4.360~GeV
provide stronger upper limits at these energy points.
The values of the upper limits measured here are listed in Table~\ref{table:up_limit}.

\begin{table}[htbp]
\caption{Upper limits on the $\EE\to \gamma\chi_{cJ}$ cross sections. }
\label{table:up_limit}
\begin{center}
\begin{tabular}{c | c  |c|| c | c |c }
\hline\hline
  $\sqrt{s}$ ($\gev$) &  $\chi_{c1}$ (pb)  & $\chi_{c2}$ (pb)
& $\sqrt{s}$ ($\gev$) &  $\chi_{c1}$ (pb)  & $\chi_{c2}$ (pb)\\\hline
3.80-3.84 & 80 & 134  & 4.68-4.72 & 8 & 14 \\\hline
3.84-3.88 & 37 & 90 &4.72-4.76 & 8& 18\\\hline
3.88-3.92 & 35 & 110&4.76-4.80 & 11& 15\\\hline
3.92-3.96 & 27 & 40 &4.80-4.84 & 9& 18\\\hline
3.96-4.00 & 12 & 21&4.84-4.88 & 15& 11\\\hline
4.00-4.04 & 34& 53 &4.88-4.92 & 11& 14\\\hline
4.04-4.08 & 29& 45&4.92-4.96 & 10& 10\\\hline
4.08-4.12 & 46& 54&4.96-5.00 & 4& 21\\\hline
4.12-4.16 & 27&53 &5.00-5.04 & 8&13\\\hline
4.16-4.20 & 10& 63 &5.04-5.08 & 13&13\\\hline
4.20-4.24 & 36& 35&5.08-5.12 & 11& 7\\\hline
4.24-4.28 & 14& 17&5.12-5.16 & 9& 7\\\hline
4.28-4.32 & 19& 38&5.16-5.20 & 5& 17\\\hline
4.32-4.36 & 16& 20&5.20-5.24 & 14& 9\\\hline
4.36-4.40 & 8& 22&5.24-5.28 & 7&6\\\hline
4.40-4.44 & 14& 34&5.28-5.32 & 6& 8\\\hline
4.44-4.48 & 11& 22&5.32-5.36 & 4& 16\\\hline
4.48-4.52 & 11&21 &5.36-5.40 & 6& 14\\\hline
4.52-4.56 & 7& 12&5.40-5.44 & 4& 10\\\hline
4.56-4.60 & 16& 13&5.44-5.48 & 8& 8\\\hline
4.60-4.64 & 6& 26&5.48-5.52 & 8& 8\\\hline
4.64-4.68 & 12&20 &5.52-5.56 & 4& 14\\\hline
\hline
\end{tabular}
\end{center}
\end{table}

We extract the transition rate of the vector charmonium
and charmoniumlike states to $\gamma\chi_{cJ}$ by fitting the
distributions in Fig.~\ref{mgchc_high_comp}.
We use a Breit-Wigner function for the signal and a first- or second-order polynomial function
for the background.
While doing the fit, the mass and total width are fixed to the
world average-values~\cite{PDG} and
$\Gamma_{ee} \times \BR({R \to \gamma\chi_{cJ}})$ is scanned 
from zero to a large number at which the probability is less
than 1.0\% of the largest value. 
Normalized probability density functions are derived from such a scan.
These  probability density functions then give the upper limits
at 90\% C.L. as
listed in Table~\ref{tab:up_ee}.
%$\Gamma_{ee}(\psi(4040)) \times \BR(\psi(4040) \to \gamma\chi_{c1}) < 3.0~\ev$,
%$\Gamma_{ee}(\psi(4040)) \times \BR(\psi(4040) \to \gamma\chi_{c2}) < 3.2~\ev$,
%$\Gamma_{ee}(\psi(4160)) \times \BR(\psi(4160) \to \gamma\chi_{c1}) < 2.4~\ev$,
%$\Gamma_{ee}(\psi(4160)) \times \BR(\psi(4160) \to \gamma\chi_{c2}) < 6.4~\ev$,
%$\Gamma_{ee}(\psi(4415)) \times \BR(\psi(4415) \to \gamma\chi_{c1}) < 0.6~\ev$,
%$\Gamma_{ee}(\psi(4415)) \times \BR(\psi(4415) \to \gamma\chi_{c2}) < 1.3~\ev$,
%$\Gamma_{ee}(Y(4260)) \times \BR(Y(4260) \to \gamma\chi_{c1}) < 1.6~\ev$,
%$\Gamma_{ee}(Y(4260)) \times \BR(Y(4260) \to \gamma\chi_{c2}) < 2.0~\ev$,
%$\Gamma_{ee}(Y(4360)) \times \BR(Y(4360) \to \gamma\chi_{c1}) < 0.6~\ev$,
% and $\Gamma_{ee}(Y(4360)) \times \BR(Y(4360) \to \gamma\chi_{c2}) < 1.2 ~\ev$.
Taking $\Gamma_{ee}[\psi(4040)]$ and
$\Gamma_{ee}[\psi(4415)]$ from the
world average-values~\cite{PDG} and $\Gamma_{ee}[\psi(4160)]$
from the BES II measurement~\cite{bes4160}, we set the
upper limits on the branching fractions for these three conventional
charmonium states
as listed in Table~\ref{tab:up_ee_r}.
Taking $\Gamma_{ee}[Y(4260)] \times \BR[Y(4260) \to \ppjpsi] = (6.4 \pm 0.8 \pm 0.6)~\ev$
or $(20.5 \pm 1.4 \pm 2.0)$~eV~\cite{ppjpsi} 
(there are two solutions for the best fit in this mode, and there are also two solutions 
in the $Y(4360)$ and $Y(4660)$ cases below)
),
$\Gamma_{ee}[Y(4360)] \times \BR[Y(4360) \to \pppsip] = (10.4  \pm 1.7 \pm 1.4)~\ev$
or $(11.8\pm 1.8\pm 1.4)$~eV~\cite{pppsp},
and $\Gamma_{ee}[Y(4660)] \times \BR[Y(4660) \to \pppsip] = (3.0  \pm 0.9 \pm 0.3)~\ev$
or $(7.6\pm 1.8 \pm 0.8)$~eV~\cite{pppsp},
we set the upper limits on the ratios of the branching fractions as shown in Table~\ref{tab:up_ee_r_2}.
%$\frac{\BR[Y(4260) \to \gamma\chi_{c1}]}{\BR[Y(4260) \to \ppjpsi]} < 0.3$ or 0.07,
%$\frac{\BR[Y(4260) \to \gamma\chi_{c2}]}{\BR[Y(4260) \to \ppjpsi]} < 0.7$ or 0.3,
%$\frac{\BR[Y(4360) \to \gamma\chi_{c1}]}{\BR[Y(4360) \to \pppsip]} < 0.06$ or 0.05,
%$\frac{\BR[Y(4360) \to \gamma\chi_{c2}]}{\BR[Y(4360) \to \pppsip]} < 0.2$ or 0.2,
%$\frac{\BR[Y(4660) \to \gamma\chi_{c1}]}{\BR[Y(4660) \to \pppsip]} < 0.2$ or 0.07
%and $\frac{\BR(Y(4660) \to \gamma\chi_{c2})}{\BR[Y(4660) \to \pppsip]} < 0.9$ or 0.3
%at 90\% C.L.
%Here the denominators are lowered by one standard deviation
%in calculating the upper limits of the ratios.
The mass and width of the vector charmonium and charmoniumlike states,
the background shape, and the fit range are varied in the fit to estimate the systematic uncertainties.
The largest upper limit from these tests is taken as the final result.
The total uncertainties from the reference processes and the systematic errors
are considered by assuming they are Gaussian errors.

\begin{table}[htbp]
\caption{Upper limits on $\Gamma_{ee} \times \BR$  at the 90\% C.L.}
\label{tab:up_ee}
\begin{tabular}{c | c | c}
\hline\hline
  &  $\chi_{c1}$~(\rm eV) &  $\chi_{c2}$~(\rm eV)\\\hline
$\Gamma_{ee}[\psi(4040)] \times \BR[\psi(4040) \to \gamma\chi_{cJ}]$ & 2.9& 4.6\\\hline
$\Gamma_{ee}[\psi(4160)] \times \BR[\psi(4160) \to \gamma\chi_{cJ}]$ & 2.2 & 6.1\\\hline
$\Gamma_{ee}[\psi(4415)] \times \BR[\psi(4415) \to \gamma\chi_{cJ}]$ & 0.47 & 2.3\\\hline
$\Gamma_{ee}[Y(4260)] \times \BR[Y(4260) \to \gamma\chi_{cJ}]$ & 1.4 & 4.0\\\hline
$\Gamma_{ee}[Y(4360)] \times \BR[Y(4360) \to \gamma\chi_{cJ}]$ & 0.57  & 1.9\\\hline
$\Gamma_{ee}[Y(4660)] \times \BR[Y(4660) \to \gamma\chi_{cJ}]$ & 0.45  & 2.1\\\hline\hline
\end{tabular}
\end{table}

\begin{table}[htbp]
\caption{Upper limits on branching fractions $\BR(R \to \gamma \chi_{cJ})$ at the 90\% C.L.}
\label{tab:up_ee_r}
\begin{tabular}{c  | c | c}
\hline\hline
Resonance  &  $\gamma\chi_{c1}~(\rm 10^{-3})$ &  $\gamma\chi_{c2}~(\rm 10^{-3})$\\\hline
$\psi(4040)$ & 3.4& 5.5\\\hline
$\psi(4160)$  & 6.1 & 16.2\\\hline
$\psi(4415)$  & 0.83  & 3.9\\\hline\hline
\end{tabular}
\end{table}

\begin{table}[htbp]
\caption{Upper limits on branching fraction ratios at the 90\% C.L. The two upper limits correspond to the two solutions
in the reference processes.}
\label{tab:up_ee_r_2}
\begin{tabular}{c  | c | c}
\hline\hline
Resonance  &  $\gamma\chi_{c1}$ &  $\gamma\chi_{c2}$\\\hline
$\frac{\BR[Y(4260) \to \gamma\chi_{cJ}]}{\BR[Y(4260) \to \ppjpsi]}$ & 0.3 or 0.07 & 0.7 or 0.2\\\hline
$\frac{\BR[Y(4360) \to \gamma\chi_{cJ}]}{\BR[Y(4360) \to \pppsip]}$ &0.06 or 0.05  &0.2 or 0.2 \\\hline
$\frac{\BR[Y(4660) \to \gamma\chi_{c1}]}{\BR[Y(4660) \to \pppsip]}$  & 0.2 or 0.07  & 0.9 or 0.3\\\hline\hline
\end{tabular}
\end{table}

The following sources of systematic uncertainties are considered
in the $\sigma(\EE \to \gamma\chi_{cJ})$ upper-limit
determination. The uncertainty in the tracking efficiency for
tracks with angles and momenta characteristic of signal events is
about 0.35\% per track~\cite{etaj} and is additive. The
uncertainty due to particle identification efficiency is 1.9\%.
The uncertainty of $\jpsi$ mass and $\chi_{cJ}$ mass requirements
are estimated using the $\psp$ sample in the same analysis and
they are found to be 1\% and 1.3\%, respectively.
The generator {\sc evtgen} is used in generating signal MC events.
In this generator, however, only one ISR photon is allowed and
the higher-order ISR effect should be estimated and corrected. 
This effect is studied by using a control sample
$\EE\to \gamma_{\rm ISR}\psp$ with $\psp$ decaying into $\eta\jpsi$.
 This process can be generated with both {\sc evtgen} and {\sc phokhara},
a generator with higher-order ISR corrections. We assume that the
correction factor obtained in this mode is the same as in the
mode under study, and 9.0\% is taken as the systematic error, corresponding to 
 the uncertainty in the difference between the measured 
 $\BR(\psp\to\gamma\chi_{cJ}\to\gamma\gamma\jpsi)$
 and the world average~\cite{PDG}.
Taking the statistical
error of the MC samples and the possible uncertainty in simulating
the angular distributions of the full decay chain
$\gamma\chi_{cJ}\to \gamma\gamma\jpsi$ into account, we quote
a total uncertainty due to the generator as 12\%.
Belle measures luminosity
with 1.4\% precision and the trigger efficiency is about 91\% with
an uncertainty of 2\%. Errors on the branching fractions of the
intermediate states are taken from Ref.~\cite{PDG} with a
systematic error of 4.5\%. Assuming that these systematic
error sources are independent, the total systematic error is 13.4\%.
The systematic uncertainty is considered in the upper limits
shown in Tables~\ref{table:up_limit}---\ref{tab:up_ee_r_2}.

In summary, using the full Belle data sample, we measure the
$\EE\to \gamma\chi_{cJ}$ process via initial state radiation. For the CM
energy between 3.80 and 5.56~$\gev$, there are no significant
$\EE\to \gamma \chi_{c1}$ and $\gamma \chi_{c2}$ signals.
The upper limits on the $\EE\to \gamma\chi_{cJ}$ production cross
sections, which range from a few pb to a few tens of pb,
are set for the first time and are listed in Table~\ref{table:up_limit}. 
We also set upper limits
on the decay rate of the vector charmonium and charmoniumlike states
to $\gamma\chi_{cJ}$. This information may help in understanding
the nature of these vector states.

%%%%%%%%%%%%%%%%%%%%%%%%%%%%%%%%%%%%%
%%%% Acknowledge add later %%%%%%%%%%
%%%%%%%%%%%%%%%%%%%%%%%%%%%%%%%%%%%%%

We thank the KEKB group for the excellent operation of the
accelerator; the KEK cryogenics group for the efficient
operation of the solenoid; and the KEK computer group,
the National Institute of Informatics, and the 
PNNL/EMSL computing group for valuable computing
and SINET4 network support.  We acknowledge support from
the Ministry of Education, Culture, Sports, Science, and
Technology (MEXT) of Japan, the Japan Society for the 
Promotion of Science (JSPS), and the Tau-Lepton Physics 
Research Center of Nagoya University; 
the Australian Research Council and the Australian 
Department of Industry, Innovation, Science and Research;
Austrian Science Fund under Grant No.~P 22742-N16 and P 26794-N20;
the National Natural Science Foundation of China under Contracts 
No.~10575109, No.~10775142, No.~10875115, No.~11175187, and  No.~11475187;
the Chinese Academy of Science Center for Excellence in Particle Physics; 
the Ministry of Education, Youth and Sports of the Czech
Republic under Contract No.~LG14034;
the Carl Zeiss Foundation, the Deutsche Forschungsgemeinschaft
and the VolkswagenStiftung;
the Department of Science and Technology of India; 
the Istituto Nazionale di Fisica Nucleare of Italy; 
National Research Foundation (NRF) of Korea Grants
No.~2011-0029457, No.~2012-0008143, No.~2012R1A1A2008330, 
No.~2013R1A1A3007772, No.~2014R1A2A2A01005286, No.~2014R1A2A2A01002734, 
No.~2014R1A1A2006456;
the Basic Research Lab program under NRF Grant No.~KRF-2011-0020333, 
No.~KRF-2011-0021196, Center for Korean J-PARC Users, No.~NRF-2013K1A3A7A06056592; 
the Brain Korea 21-Plus program and the Global Science Experimental Data 
Hub Center of the Korea Institute of Science and Technology Information;
the Polish Ministry of Science and Higher Education and 
the National Science Center;
the Ministry of Education and Science of the Russian Federation and
the Russian Foundation for Basic Research;
the Slovenian Research Agency;
the Basque Foundation for Science (IKERBASQUE) and 
the Euskal Herriko Unibertsitatea (UPV/EHU) under program UFI 11/55 (Spain);
the Swiss National Science Foundation; the National Science Council
and the Ministry of Education of Taiwan; and the U.S.\
Department of Energy and the National Science Foundation.
This work is supported by a Grant-in-Aid from MEXT for 
Science Research in a Priority Area (``New Development of 
Flavor Physics'') and from JSPS for Creative Scientific 
Research (``Evolution of Tau-lepton Physics'').


\begin{thebibliography}{**}

\bibitem{belley} C.~Z.~Yuan {\it et al.} (Belle Collaboration),
\Journal\PRL{99}{182004}{2007}.

\bibitem{babay4260} B.~Aubert {\it et al.} (BaBar Collaboration),
\Journal\PRL{95}{142001}{2005}; J. P. Lees {\it et al.} (BaBar
Collaboration), \Journal\PRD{86}{051102}{2012}.

\bibitem{pppsp} X.~L. Wang {\it et al.} (Belle Collaboration),
\Journal\PRL{99}{142002}{2007}.

\bibitem{babay4324} J.~P. Lees {\it et al.} (BaBar Collaboration),
\Journal\PRD{89}{111103}{2014}.

\bibitem{barnes} S.~Godfrey and N. Isgur,
\Journal\PRD{32}{189}{1985}; T.~Barnes, S. Godfrey and E. S. Swanson,
\Journal\PRD{72}{054026}{2005}; G.~J.~Ding, J. J. Zhu and M. L. Yan,
\Journal\PRD{77}{014033}{2008}.

\bibitem{review} For a review, see N.~Brambilla {\it et al.},
\Journal\EPJC{71}{1534}{2011}.

\bibitem{cleo_br} T.~E.~Coan {\it et al.} (CLEO Collaboration),
\Journal\PRL{96}{162003}{2006}.

\bibitem{bes3} M. Ablikim {\it et al.} (BESIII Collaboration),
Chin. Phys. C {\bf 39}, 041001 (2015).

%\bibitem{Belle} A.~Abashian {\it et al.} (Belle Collaboration),
%\Journal\NIMA{479}{117}{2002}.

\bibitem{Belle}A.~Abashian {\it et al.} (Belle Collaboration), Nucl. Instrum. Methods
 Phys. Res. Sect. A {\bf 479}, 117 (2002); also see detector section in
 J.Brodzicka {\it et al.}, Prog. Theor. Exp. Phys. {\bf 2012}, 04D001 (2012).

%\bibitem{KEKB} S.~Kurokawa and E.~Kikutani,
%\Journal\NIMA{499}{1}{2003} and other papers included in this
%volume.

\bibitem{KEKB}S.~Kurokawa and E.~Kikutani, Nucl. Instrum. Methods Phys. Res. Sect.
 A {\bf 499}, 1 (2003), and other papers included in this Volume;
 T.Abe {\it et al.}, Prog. Theor. Exp. Phys. {\bf 2013}, 03A001 (2013) and following
 articles up to 03A011.

\bibitem{etaj} X.~L.~Wang {\it et al.} (Belle Collaboration),
\Journal\PRD{87}{051101}{2013}.

\bibitem{EvtGen} D.~J.~Lange, {Nucl. Instrum. Methods Phys. Res. 
  Sect.} A {\bf 462}, 152 (2001).

\bibitem{phokhara} G.~Rodrigo {\it et al.},
\Journal\EPJC{24}{71}{2002};
S.~Actis {\it et al.}, \Journal\EPJC{66}{585}{2010}.

\bibitem{geant4}
R. Brun {\it et al.}, GEANT3.21, CERN Report DD/EE/84-1 (1984). 

\bibitem{Rlike}E. Nakano,
\Journal\NIMA{494}{402}{2002}.

\bibitem{MUID} A. Abashian {\it et al.},
% K. Abe, K. Abe, P.K. Behera, F. Handa,
%T. 8Iijima, Y. Inoue, H. Miyake, T. Nagamine, E. Nakano, S.
%Narita, L. Piilonen, S. Schrenk, Y. Teramoto, K. Trabelsi, J. G.
%Wang, M. Yamaga, A. Yamaguchi, Y. Yusa,
\Journal\NIMA{491}{69}{2002}.


%\bibitem{PDG} J.~Beringer {\it et al.} (Particle Data Group),
%\Journal\PRD{86}{010001}{2012} and 2013 partial update for the 2014 edition.

\bibitem{ISR}E.~A.~Kuraev and V.~S.~Fadin, Yad. Fiz. {\bf 41}, 733 (1985)
[Sov. J. Nucl. Phys. {\bf 41}, 466 (1985)].

\bibitem{PDG} K. A. Olive {\it et al.} (Particle Data Group), 
Chin. Phys. C, {\bf 38}, 090001 (2014). 

%\bibitem{cleo} H.~Mendez {\it et al.} (CLEO Collaboration),
%\Journal\PRD{78}{011102}{2008}.

\bibitem{ppjpsi}Z.~Q. Liu {\it et al.} (Belle Collaboration),
\Journal\PRL{110}{252002}{2013}.



\bibitem{bes4160}M. Ablikim {\it et al.} (BES Collaboration),
\Journal\PLB{660}{315}{2007}.


%%%%%%%%%%%%%%%%%%%%%%%%%%%%
\end{thebibliography}
\end{document}